# Automated Identification and Classification of Stereochemistry: Chirality and Double Bond Stereoisomerism

Ana L. Teixeira,[*,†,‡] João P. Leal[‡, §] and Andre O. Falcao[†]

[†] LaSIGE, Departamento de Informática, Faculdade de Ciências, Universidade de Lisboa, 1749-016 Lisboa, Portugal; [‡] Centro de Química e Bioquímica, Faculdade de Ciências, Universidade de Lisboa, 1749-016 Lisboa, Portugal; [§] Unidade de Ciências Químicas e Radiofarmacêuticas, Instituto Tecnológico e Nuclear, Instituto Superior Técnico, Universidade Técnica de Lisboa, 2686-953 Sacavém, Portugal.

## Abstract

Stereoisomers have the same molecular formula and the same atom connectivity and their existence can be related to the presence of different three-dimensional (3D) arrangements. Stereoisomerism is of great importance in many different fields since the molecular properties and biological effects of the stereoisomers are often significantly different. Most drugs for example, are often composed of a single stereoisomer of a compound, and while one of them may have therapeutic effects on the body, another may be toxic. A challenging task is the automatic detection of stereoisomers using line input specifications such as SMILES or InChI since it requires information about group theory (to distinguish stereoisomers using mathematical information about its symmetry), topology and geometry of the molecule.

There are several software packages that include modules to handle stereochemistry, especially the ones to name a chemical structure and/or view, edit and generate chemical structure diagrams. However, there is a lack of software capable of automatically analyzing a molecule represented as a graph and generate a classification of the type of isomerism present in a given atom or bond. Considering the importance of stereoisomerism when comparing chemical structures, this report describes a computer program for analyzing and processing steric information contained in a chemical structure represented as a molecular graph and providing as output a binary classification of the isomer type based on the recommended conventions. Due to the complexity of the underlying issue, specification of stereochemical information is currently limited to explicit stereochemistry and to the two most common types of stereochemistry caused by asymmetry around carbon atoms: chiral atom and double bond.

* Corresponding author.
*E-mail address:* ateixeira@lasige.di.fc.ul.pt

# 1. Introduction

One of the major tasks in cheminformatics is to represent chemical structures and to transfer the various types of representation into computer-readable formats [1]. The complexity of the chemical information and the difficulties representing and analyzing it constitute a considerable challenge in informatics. Molecules consist of atoms held together by covalent chemical bonds. Furthermore, molecules can be transformed into other molecules by the process of chemical reactions. Therefore, chemical information not only comprises text and numbers but also has to characterize chemical compounds with their special properties, geometries, interactions and reactions [1-4]. A particular issue which arises in the representation of chemical structures is to determinate how much of the above mentioned information to include [1, 2, 4]. The purpose of machine-readable structure representations is to mine the molecular information and make it suitable to perform common operations on molecules such as storage/retrieval, identity, substructure/superstructure relationships, similarity and multivariate relationships [2, 5]. Chemical structures are normally stored in a computer as molecular graphs. In a molecular graph (usually non-directed and labeled), the nodes correspond to the atoms and the edges to the bonds. Its vertices and edges are labeled with the kinds of the corresponding atoms or bond types, respectively [1, 2, 6]. The molecular graphs, for computer processing, are transformed (manually or automatically) into linear strings of characters (for example, SMILES [7] or InChI [8]) or into two-dimensional matrices (for example connection tables or adjacency matrices [9]) listing all of the atoms (nodes) with their mutual interconnections (bonds or edges) [1, 4, 6]. Figure 1 exemplifies different representations for a simple molecule, ethanol.

| Structure | Molecular Representations | |
|---|---|---|
| | *Linear Notations* | |
| | **SMILES** | CCO |
| | **InChI** | InChI=1S/C2H6O/c1-2-3/h3H,2H2,1H3 |
| | *2D matrices* | |
| (structure of ethanol: $^1H_3C$—$CH_2$—$OH^3$, bond labeled 2) | **Connection Table** (excluding hydrogen (H)) | 3 2 0 0 0 0 0 0 0 0999 V2000<br>1.7321  -0.0000   0.0000 C<br>0.8660  -0.5000   0.0000 C<br>0.0000  -0.0000   0.0000 O<br>**1 2 1** 0 0 0 0<br>**2 3 1** 0 0 0 0 |
| | **Adjacency Matrix** (excluding hydrogen (H)) | atom \| 1 \| 2 \| 3<br>1 \| 6 \| 1 \| 0<br>2 \| 1 \| 6 \| 1<br>3 \| 0 \| 1 \| 8 |

**Figure 1. Different representations for the molecule ethanol ($C_2H_6O$).**

In this context, a graph represents the topology of a molecule, i.e. the way the nodes (or atoms) are connected and is less suitable for modeling those properties that are determined by molecular geometry, conformation or stereochemistry. Thus a given graph may be drawn in many different ways and may not obviously correspond to a "standard" chemical. The complexity of chemical systems is considerably reduced once some aspects are neglected, when they are modeled as graphs [1, 4, 6]. Comparing molecular graphs allows the distinction between structural isomers (compounds with the same molecular formula but non-isomorphic graph), moreover, typically it does not contain information about the three-dimensional (3D) arrangement of the molecule [1, 4, 6, 9, 10]. Figure 2 exemplifies two different compounds, a) (R)-2-amino-3-(4-hydroxyphenyl) propanoic acid and b) (S)-2-amino-3-(4-hydroxyphenyl) propanoic acid with identical connectivity (Figure 2 - c), i.e. a) and b) have the same topology, but different topography.

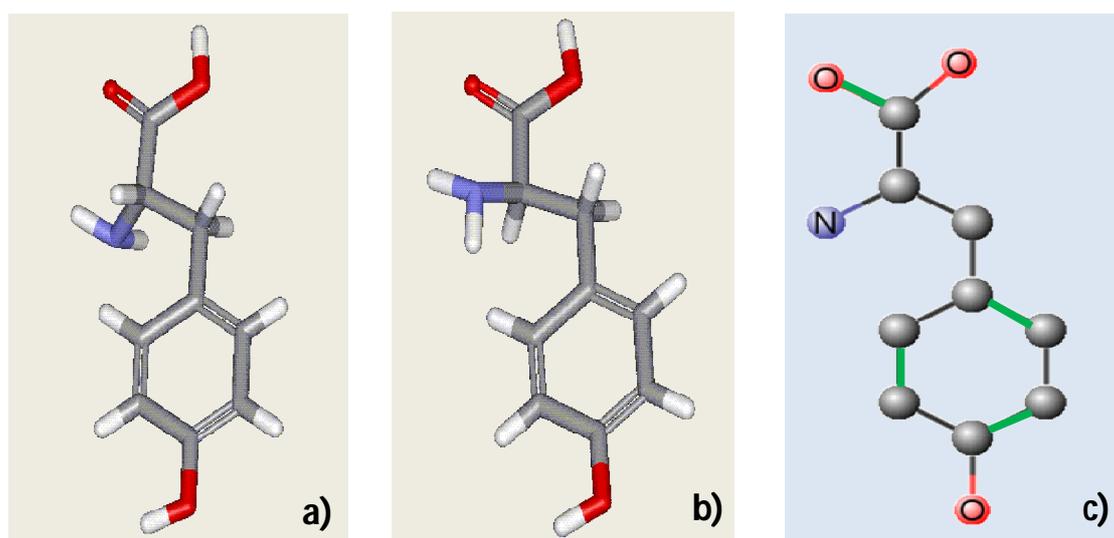

**Figure 2.** 3-D representation of **a)** (R)-2-amino-3-(4-hydroxyphenyl) propanoic acid and **b)** (S)-2-amino-3-(4-hydroxyphenyl) propanoic acid. **c)** molecular graph representation of 2-amino-3-(4-hydroxyphenyl) propanoic acid (excluding hydrogen atoms).

The fact that molecules are 3D objects leads to situations where molecules with the same set of atoms and bonds can nonetheless be different as a result of their bonds having different arrangement in space (stereoisomers) [10-13]. Consequently, the configuration of a molecule defines the position of groups around one or more non-rotating bonds (as double bonds in the case of geometric isomerism) or around a chiral center. This inability of a double bond to rotate is known as hindered rotation, transpiring in the formation of the cis (Z) isomer and the trans (E) isomer. Such pairs of geometric isomers are called double-bond diastereomers. The spatial distances in diastereomeric pairs are not all identical which results in different chemical and physical properties between diastereomeric molecules [10, 12]. The term chirality refers to the handedness of an object that is not identical with its mirror image and for that reason it cannot be superimposed. The mirror-image configurations of the chiral molecule define a pair of enantiomers with identical chemical and physical properties, with the exception of their interactions with other stereoisomers and polarized light [10, 12]. Stereoisomerism is of great importance in many different fields since it often has dramatic consequences in molecular properties such as pharmacological activity, odor, environmental impact, physical properties, or chemical reactivity, particularly in the interaction of stereoisomers. For example, the effectiveness and toxicology of drugs depends on enantiomeric selectivity and purity [14, 15].

There are several software packages that include modules to handle stereochemistry, especially the ones to name a chemical structure and/or view, edit and generate chemical structure diagrams such as Modular Chemical Descriptor Language (MCDL)[16], ChemBioDraw[17], Marvin[18], ACD/Chemsketch-[19], MOLGEN[20], among others. Another classical problem in chemoinformatics, related to this subject, is the enumeration of all possible chemical graphs for a certain structure. Several algorithms and commercially available software packages [21, 22] have been developed for the efficient enumeration of all stereoisomers of a chemical graph using different strategies[23] such as Corina[24], Stereoplex (module integrated in the software package Concord)[25], OMEGA[26], Molecular Operating Environment (MOE)[27], Catalyst[28], ChemAxon Calculator Plugins[29], FROG[30, 31], among others. Computer programs to assist organic synthesis planning have also integrated modules to recognize different stereoisomers, such as CHIRON[32] and LHASA[33]. Several authors have also integrated different stereo descriptors applied to quantitative structure-activity relationship (QSAR) problems [34-39].

The automatic detection of stereoisomers is a challenging task since it requires group theory (study the algebraic structures to distinguish stereoisomers using mathematical information about its symmetry), topology and geometry of a molecule [16]. Considering the importance of stereoisomers when comparing chemical structures, this report describes a computer program for analyzing and processing steric information contained in a chemical structure represented as a molecular graph. Most of the existing software packages are proprietary and/or commercial, and they do not allow batch processing of molecules encoded as line representations such as SMILES or InChI. This work arose from the need to have a computer program providing as output a binary classification of the isomer type according to Cahn-Ingold-Prelog (CIP) system[40] using as input widely used line notations such as SMILES or InChI, as it will be explained below. Due to the complexity of the underlying issue, specification of stereochemical information is currently limited to explicit stereochemistry and to the two most common types, stereochemistry of a chiral carbon atom and stereochemistry of a double bond.

## 2. Identification and Specification of Chirality

A simple chiral molecule has four unique chemical groups arranged around a tetrahedral carbon atom. Chiral carbon atoms are the most common source of molecular asymmetry and are found in many optically active biological compounds. If *n* asymmetric carbon atoms are present, then it is possible to generate $2^n$ isomers (stereoisomers) [12]. Chiral configurations were previously designated as D and L by Fischer in 1891 (using D-glyceraldehyde as a reference) or recently as R and S according to the R-S notation defined by Cahn in 1964 [41, 42]. In order to establish the configuration, the four groups surrounding the central carbon atom (stereocenter) are ranked according to a priority sequence, determined by a number of sequence rules. The official systematic nomenclature to determine such rules was proposed in 1966 by Cahn, Ingold and Prelog (CIP rules) [40] and later extended[43]. The CIP rules play a double role: first, they allow determining whether the considered atom is really asymmetric, and second they rank the ligands connected to the stereocenter producing a pre-defined priority. The CIP rules to define priorities to ligands can be simplified as follows [40, 43]:

**Rule 1.** Considering the atoms directly attached to the stereocenter, the higher the atomic number the higher priority;

**Rule 2.** If there are two or more atoms with the same atomic number, the method of determination is to perform a similar comparison of the atomic numbers of the next atoms in the ligands until a point of difference is found;

**Rule 3.** If two atoms have the same atomic number but different mass number (isotopes), the atom with higher mass number comes first;

**Rule 4.** Double or triple bonds are counted as if they were split into two or three single bonds, respectively;

**Rule 5.** For compounds where only configurational differences between ligands are detected, the following rules apply:

**a.** When two ligands differ only in that one has an atom of higher rank in a cis-position and the other in a trans-position, then preference is given to the former. Cis and trans double bonds rank higher than nonstereogenic double bonds (cis > trans > nonstereogenic) [44].

**b.** Chiral stereogenic units precede pseudoasymmetric stereogenic units and these precede nonstereogenic units.

**c.** When considering stereogenic units, if two ligands have different descriptor pairs, then the one with the first *like* descriptor pair has priority over the one with a corresponding *unlike* descriptor pair.

**d.** r precedes s (chirality of a pseudoasymmetric carbon).

**e.** R precedes S.

Rule 5 is used to compare combinations of chiral units, however, it can lead to ambiguity in the evaluation of pairs of ligands [45-47]. Although new methodologies (e.g [45-47]) have been purposed to overcome this ambiguity, this rule will not be considered in the present version.

The lowest priority ligand determined using the CIP rules is then selected as the reference ligand and the molecule should be oriented pointing this group to the back. The line connecting the chiral centre and the lowest priority ligand determines the reference plane. If the path from the higher to the lower priority of the three other ligands goes in the clockwise (right) direction then the chiral centre is described as R (from the Latin *rectus* for right) otherwise as S (from the Latin *sinister* for left). As an example, considering 2-amino-3-mercaptopropanoic acid (amino acid also known as cysteine with the chemical formula

*HO₂CCH(NH₂)CH₂SH*) shown in Figure 3, the α-carbon (marked with an asterisk) is chiral, and thus this molecule has two mirror-image structures (**a)** and **b)**) [48].

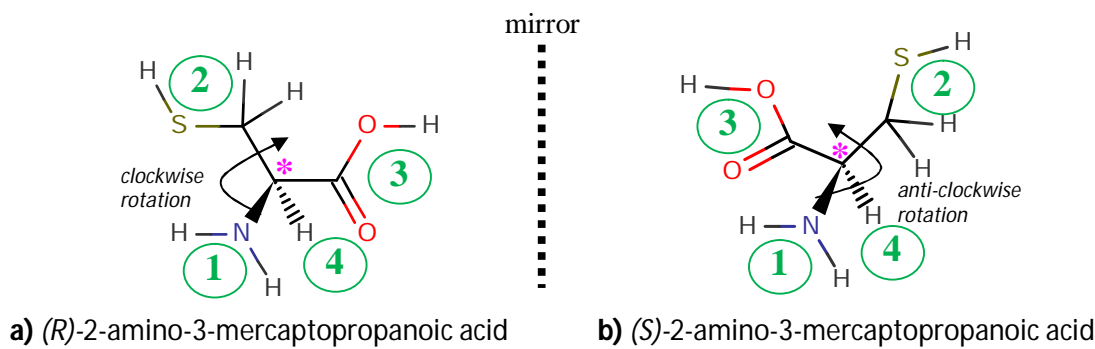

**a)** *(R)*-2-amino-3-mercaptopropanoic acid   **b)** *(S)*-2-amino-3-mercaptopropanoic acid

**Figure 3. The two enantiomers (mirror-image structures) of the compound 2-amino-3-mercaptopropanoic acid (cysteine).** The chiral carbon is marked with an asterisk and the priorities of the substituents indicated inside circles. **a)** (R)-2-amino-3-mercaptopropanoic acid: the sense of the rotation of the groups with highest priorities 1 to 3 is clockwise hence the **R** assignment. **b)** (S)-2-amino-3-mercaptopropanoic acid: the sense of the rotation of the groups with highest priorities 1 to 3 is anti-clockwise hence the **S** assignment.

The representations of 2-amino-3-mercaptopropanoic acid enantiomers are shown in Figure 3 with the priorities of the substituents indicated. The priorities are assigned based on atomic number (Z) of the atoms that are directly attached to the stereocenter and in situations where two or more substituents have the same atomic number, one should proceed along the respective substituent chain until a point of difference is found: nitrogen (Z = 7) > carbon (Z = 6) - sulfur (Z = 16 second level)  > carbon (Z = 6) - oxygen (Z = 8 second level) > hydrogen (Z = 1). Allowing hydrogen (lowest priority) to point away from the viewer, in the case of the structure represented in Figure 3 **a)** the sense of the rotation of the groups 1 to 3 is clockwise (to the right) hence the **R** assignment while in the case of the structure represented in Figure 3 **b)** the sense of the rotation of the groups 1 to 3 is anti-clockwise (to the left) hence the **S** assignment.

Chirality can be expressed in the SMILES notation. Depending on the clockwise or anti-clockwise ordering of the atoms around the stereocenter it can be identified as "@@" or "@", respectively. The sequence of the atoms in the SMILES solely depends on the order of writing, and it is independent of the priorities of the atoms, therefore there is not a direct correspondence with the R-S system [7, 49, 50]. For example, in the SMILES for (R)-2-amino-2-fluoropropanoic acid - N[C@](C)(F)C(=O)O (equivalent to F[C@@](N)(C(O)=O)C, C[C@@](N)(F)C(O)=O, among others) - looking from the atom N to the chiral center, the other ligands (methyl (CH₃), fluorine (F) and carboxy (COOH)) appear in anti-clockwise order as it is displayed in Figure 4.

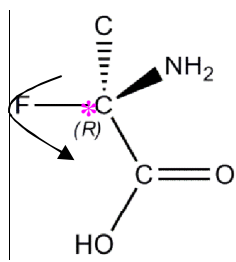

**Figure 4. The representation of the compound (R)-2-amino-2-fluoropropanoic acid using the SMILES N[C@](C)(F)C(=O)O.** In this representation looking from the atom N to the chiral center (marked with an asterisk), the other substituents (methyl (CH3), fluorine (F) and carboxy (COOH)) appear in anti-clockwise order which does not have a direct correspondence with the R-S system since the priorities of the substituents define a clockwise sense of the rotation hence the **R** assignment.

Chirality can also be expressed in the InChI (IUPAC International Chemical Identifier) notation. InChI is a string of characters capable of uniquely representing a molecule and is derived solely from a structural representation of that molecule, structured to be independent of the way that the structure was drawn. Every InChI starts with the string "InChI=" followed by the version number and by the letter S for standard InChIs. The remaining information is structured as a sequence of layers and sub-layers separated by the delimiter "/", with each layer providing one specific type of information (indicated by a prefix letter). In the case of tetrahedral stereochemistry (sub-layer of the stereochemical layer) the prefix letter is a "t" and its information is derived from 'in-out' wedge bond types or $x$, $y$, $z$ coordinates. Only explicit stereo information is included and, as in SMILES, there is no simple relation between InChI and R/S configurations of stereocenters since it does not use CIP rules and calculates parities from its own canonical numbers of atoms. The parity is '+' if the canonical numbers of neighbors increase clockwise when observed from the atom that has the smallest canonical number and it is '-' for the inverse case [8].

For example, the InChI for (R)-2-amino-2-fluoropropanoic acid is InChI=1/C3H6FNO2/c1-3(4,5)2(6)7/h5H2,1H3,(H,6,7)/**t3-**/m0/s1/i1-12,2-12,3-12,4-19,5-14,6-16,7-16     (tetrahedral stereochemistry sub-layer is marked bold) looking from the ligand with the lowest canonical number, methyl (CH$_3$) (canonical number: 1), to the chiral center (canonical number: 3), the other ligands (carboxy (COOH) (canonical number: 2), fluorine (F) (canonical number: 4) and  nitrogen (N) (canonical number: 5)) appear in anti-clockwise order.

## 2.1 Algorithm to determine chirality

The algorithm that determines the chirality of a chemical structure based on widely used linear notations as input such as SMILES or InChI was implemented in python (version 2.6) and uses OpenBabel-Pybel libraries (version 2.3.1) [51-53] which in turn use joelib2[54] for processing chemical structures. To determine the chirality according to the R-S notation several steps are needed, namely: (1) indentify the stereocenters, (2) number the atoms in the molecule skeleton, (3) assign the priority of each ligand according to the CIP rules, (4) map the ligands into the skeleton groups which results in a permutation and finally (5) determine the parity of the permutation which allows the classification of the stereocenter(s) in R or S.

The basic steps for describing a stereoisomer implemented in the algorithm are exemplified in Figure 5 for the compound (S)-1-amino-1-bromoethanol and detailed below.

| List of SMILES representing the same structure | Canonical SMILES | Numbering the atoms in the skeleton (canonical order) | Assigning the priority to each group attached to the stereocenter (CIP rules) | Mapping the ligands into the skeleton groups (permutation) | Determining the classification R-S |
|---|---|---|---|---|---|
| N[C@](Br)(O)C<br>Br[C@](O)(N)C<br>O[C@](Br)(C)N<br>Br[C@](C)(O)N<br>C[C@](Br)(N)O<br>Br[C@](N)(C)O<br>C[C@@](Br)(O)N<br>Br[C@@](N)(O)C<br>.... | C[C@@](Br)(O)N | (structure with Br 1, NH$_2$ 3, HO 2, CH$_3$ 0) | (structure with Br 0, NH$_2$ 2, HO 1, CH$_3$ 3) | [CH$_3$, Br, OH, NH$_2$]<br>↓ ↓ ↓ ↓<br>[Br, OH, NH2, CH3]<br><br>Permutation = [1,2,3,0] | Initial sense of rotation (1), odd permutation (-1), (1)*(-1) = -1 => **S** |
| Canonicalization Method ⇨ | Chiral center (@@) is clockwise **(1)** ⇨ | [0,1,2,3]<br>[CH$_3$, Br, OH, NH$_2$] ⇨ | [0,1,2,3]<br>[Br, OH, NH$_2$, CH$_3$]<br>Z=[35, 8, 7, 6] ⇨ | Odd Permutation **(-1)** ⇨ | **(S)-1-amino-1-bromoethanol** |

**Figure 5. Basic steps for classifying a stereoisomer of the compound 1-amino-1-bromoethanol using the described algorithm**: the input notation is converted in a canonical SMILES, the initial sense of rotation is determined, at the stereocenter the atoms are then separated into the skeleton and its ligands and skeleton both are numbered independently in order to determine the permutation, finally the stereocenter can be classified as R or S combining the parity of the permutation and the initial sense of rotation.

The process of determining chirality of a molecule can be detailed as follows:

### a. Identifying the chiral centers

This computer program processes only molecules with explicit chirality, i.e. carbon atoms with four different groups attached which causes a lack of symmetry and is explicitly described in SMILES with @ or @@. The identification of a chiral center is based on the openbabel's boolean function `IsChiral`. After identifying the presence of one or more chiral centers, one must classify these chiral centers as clockwise or anti-clockwise. For that purpose, the arbitrary numbering of the atoms has to be canonicalized. The representation of a chemical structure by a linear notation is neither unambiguous (since often more than one structure can be produced from one representation) nor unique (since often several representations can be derived for a structure). A molecule may be, in principle, denoted with n! (n = number of atoms) different SMILES with different numbering of the atoms. The advantage of canonicalization is that it will represent a molecule in a unique order of appearance in the structure dealing with the possibility of symmetry equivalent stereocenters. It is important to note that these canonical SMILES are usually not transferable between implementations because they use different canonicalization algorithms. In this case, the canonicalization method in use is the one implemented by openbabel which is based on the widely used Morgan algorithm[55] but with various improvements [51, 52]. The main part of the Morgan algorithm is the iterative calculation of connectivity indices to enable differentiation of the atoms. At this point, a chiral center is classified as clockwise (1) if the canonical SMILES contains "@@" or anti-clockwise (-1) if the canonical SMILES contains

"@". For example, the compound (S)-1-amino-1-bromoethanol can be represented by the canonical SMILES C[C@@](Br)(O)N, therefore the chiral center has a clockwise ("@@") sense of rotation and is classified as **1** (Figure 5).

### b. Numbering the atoms in the molecule skeleton

Since the atom numbering in canonical SMILES represents the order of appearance in the structure, it can be used to number the atoms in the molecule skeleton attached to the stereocenter starting from the lowest to the highest number and renumbering them to [0, 1, 2, 3]. For the example described in Figure 5, the canonical order (C[C@@](Br)(O)N) determines that 0 corresponds to the group $CH_3$, 1 corresponds to the group Br, 2 corresponds to the group OH and 3 corresponds to the group $NH_2$.

### c. Assigning the priority to each group attached to the stereocenter

The ligands attached to the stereocenter have to be numbered independently from step **b** and according to the CIP rules. In order to establish the priority sequence, one must look at the atomic number of the atoms that are bonded directly to the stereocenter, arranging them in decreasing order of atomic number or, optionally, atomic mass to take into account isotopes. If two or more atoms bonded directly to the stereocenter have the same atomic number, then look for the first point of difference in the branch by moving out one atom (level) at a time, again comparing atomic numbers/atomic mass. This recursive method also takes into account double or triple bonds (fourth CIP rule), since higher bond orders have a smaller number of hydrogen atoms and *vice-versa*. For the example described in Figure 5, the group with highest priority is Br since its atomic number is 35, followed by O since its atomic number is 8, N since its atomic number is 7, and finally C since its atomic number is 6.

### d. Mapping the ligands into the skeleton groups (permutation)

The four ligands can be arranged around the stereocenter according to their priority (0-3) in 24 different ways. The 24 permutations of different priorities of the ligands can be grouped into two classes because of the symmetry of a tetrahedron: even and odd. Interchanging ligands starting with the initial order determined in **b**, until they are in descending order of priority according to the CIP rules determined in **c**, leads to the classification of the stereoisomer depending on the number of permutations. The permutation representation allows the comparison of the skeleton numbering and the sequence of the ligands.

### e. Determining the classification R-S using the parity of the permutation

In order to determine the parity of the permutation, i.e., whether a given permutation is even (even number of 2 ligands swaps) or odd (odd number of 2 ligands swaps), one of the most efficient methods is to write the permutation as a product of disjoint cycles. A cycle is a permutation which maps a finite subset $\{x_1, x_2, ..., x_n\}$ by $x_1 \to x_2 \to ... \to x_n \to x_1$ and will be denoted ($x_1$ $x_2$ ... $x_n$). For example, a cycle of length 2 is called a transposition and it represents a permutation that swaps two elements and leaves everything else fixed. The cycles of length 1 can be omitted from the cycle list, since they represent a ligand which does not move [56, 57]. A general procedure to calculate the disjoint cycles starts with an element and follows its correspondences under the permutation until it goes back to the starting element again. Until all the elements have been visited, this procedure should be repeated. The permutation is odd if this factorization contains an odd number of even-length cycles, for the remaining situations the permutation is even. In the case of determining the chirality of a molecule (permutation with length 4) the followed approach classifies even permutations as **1** and odd permutations as **-1**. According to the classification into clockwise of anti-clockwise determined in step **a** and the parity of the permutation it is now possible to classify the molecule in one of the two classes: R or S. For that purpose one must multiply the indexes of the initial sense of rotation and the parity of the permutation, if the result is **1** the stereocenter is classified as **R**, otherwise if the result is -**1** the stereocenter is classified as **S**.

For the example described in Figure 5, the initial atom order ([$CH_3$, Br, OH, $NH_2$]) go through 3 modifications to be ordered according to the CIP rules ([Br, OH, $NH_2$, $CH_3$]): 0**->**1**->**2**->**3. This permutation can be factorized as (0123) and since it is 1 cycle (odd number of cycles) of 4 elements (even length), the permutation is odd (-1). The initial sense of rotation is 1, multiplied by an odd permutation (-1) obtains -1 and the stereocenter is classified as S.

## 2.2 Detailed example of the steps of execution of the chirality classification program

In order to analyze in a more detailed way the execution of the developed program, another example, R-2-amino-3-mercaptopropanoic acid (already described in Figure 3 and represented again in Figure 6), will be described together with the program's outcomes in each step.

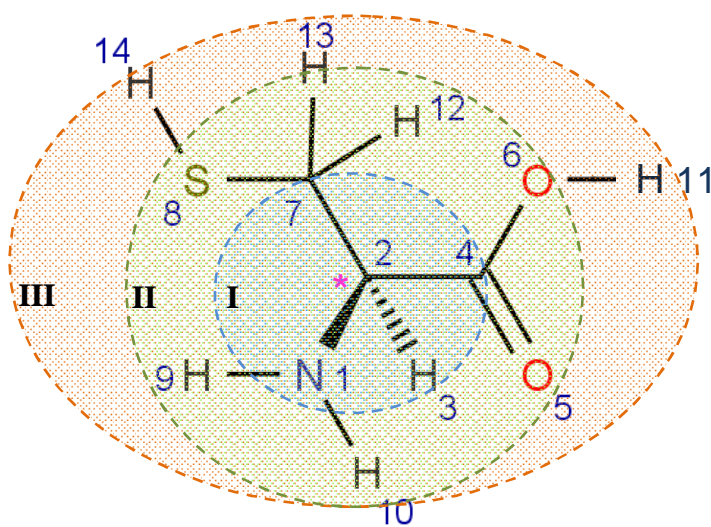

**Figure 6. The compound R-2-amino-3-mercaptopropanoic acid.** The chiral carbon is marked with an asterisk (canonical number, 2) and the canonical numbers of each atom are indicated near each atom symbol. To apply outward exploration of groups, starting from the chiral atom, the atoms are divided in levels, which are indicated in the figure with colored shells: **I)** level 1, marked with a blue shell, containing the atoms with the canonical numbers 1, 3, 4 and 7; **II)** level 2, marked with a green shell, containing the atoms with the canonical numbers 5, 6, 8, 9, 10, 12 and 13; **III)** level 3, marked with a orange shell, containing the atoms with the canonical numbers 11 and 14. The atomic numbers (Z) of the atoms represented in the figure are: sulfur (S) (Z= 16), oxygen (O) (Z = 8), nitrogen (N) (Z = 7), carbon (C) (Z = 6), hydrogen (H) (Z = 1).

The canonicalization of this structure generates the following SMILES, N[C@H](C(=O)O)CS, which indicates that this molecule has a chiral center ('@') with an anti-clockwise (**-1**) sense of rotation. Figure 6 represents the structure of R-2-amino-3-mercaptopropanoic acid, with the atoms numbered according to the canonical order. Thus, a simple 5-steps process is followed to determine its chiral orientation:

1. A recursive process is used to retrieve the atomic numbers of each atom directly attached to the stereocenter and each branch is traversed recursively. Four lists are created, each of them starting with the atomic number (Z) of the atom directly attached to stereocenter (N (index = 1, Z = 7), H (index = 3, Z = 1), C (index 4, Z = 6), C (index 7, Z = 6)), followed by lists representing a sequential topological distance to the stereocenter (levels) with two-atoms tuples (canonical indexes) indicating the connectivity of each branch:

    **[7, [(1, 9), (1, 10)]]**
    **[1]**
    **[6, [(4, 5), (4, 6)], [(6, 11)]]**
    **[6, [(7, 12), (7, 13), (7, 8)], [(8, 14)]]**

2. The atoms of each branch are sorted in descending order of their topological distances to the stereocenter (levels) and inside each level in descending order according the atomic numbers. Each branch is represented by a tuple and the connectivity tuples are replaced by lists of atomic numbers (ordered in descending order) of the atoms in each level and the last position of each tuple represents the branch number according to the canonical order (from 0 to 3):

    **([[7], [1, 1]], 0)**
    **([[1]], 1)**
    **([[6], [8, 8], [1]], 2)**
    **([[6], [16, 1, 1], [1]], 3)**

3. The lists are then compared between themselves to retrieve their priority order, as per the CIP rules. This involves solely comparing each list in the four tuples to the others to sort the branches according to their atom content:

    **([[7], [1, 1]], 0)**
    **([[6], [16, 1, 1], [1]], 3)**
    **([[6], [8, 8], [1]], 2)**
    **([[1]], 1)**

4. The permutation resulting from the last step, coupled with the canonical order of the branches (last digit of the tuple) can be used to compare the canonical order with the priority order according to the CIP rules using its parity:

    Permutation: **[0, 3, 2, 1]**, which can be represented as product of disjoint cycles as follows: (13), this factorization contains and odd number (1 cycle) of even-length (2) cycles, consequently the parity of this permutation is odd (- **1**).

5. Finally, it is possible to determine the classification in R-S using the initial sense of rotation and the parity of the permutation:

    Anti-clockwise sense of rotation (**-1**) * odd permutation (**-1**) = **1,** which corresponds to an **R** stereocenter.

The development of this computer program was guided and tested by calculating chirality to a set of 210 molecules with this type of stereochemistry. The worst-case time complexity of an algorithm gives an idea of the number of operations needed to solve it given any input of size *n* [58]. In the case of the given computer program, the most complex step is the recursive search for the ligands attached to each stereocenter, in the worst-case scenario for each stereocenter it goes through all atoms in the molecule. The complexity for the worst-case scenario (using the big O notation) of the given step, based on well-established algorithms and their known complexity [59], is *O(C\*A\*log(A))*, where *C* is the number of chiral atoms and *A* is the number of atoms in the molecule.

## 3. Identification and Specification of Double Bond Stereoisomerism

The geometrical (cis-trans) stereoisomerism arises when substituents are arranged differently in space due to restricted rotation of a double bond in a molecule [10, 12]. The carbon-carbon double bond is formed between two $sp^2$ hybridized carbons, and consists of two occupied molecular orbitals, sigma and pi. Rotation of the end groups of a double bond relative to each other destroys the p-orbital overlap that creates the pi orbital or bond. The pi bond is an high energy bond and therefore, induces a resistance to rotation which stabilizes the planar configuration [14, 15]. In order to be considered a stereo bond, the substituents on either end of the double bond have to be different. To each substituent on a double bond is assigned a priority based on the Cahn-Ingold-Prelog (CIP) priority rules [40, 43] described above and then classified according to the latest IUPAC recommendation [60], the E-Z convention [10, 12]. This convention defines that the substituents with higher priority on each side of the double bond define the reference frame. If the two substituents of higher priority are on opposite sides of the double bond (*trans* arrangement), then the **E** (from the German *entgegen* for opposite) configuration is assigned to the bond. If the two groups of higher priority are on the same side of the double bond (*cis* arrangement), then the **Z** (from the German *zusammen* for together) configuration is assigned to it [10, 12].
As an example, considering the two isomers of 2-chloro-2-butene shown in Figure 9, the stereochemistry of this alkene can be determined by comparing the relative positions of the highest priority groups on the double bond carbons.

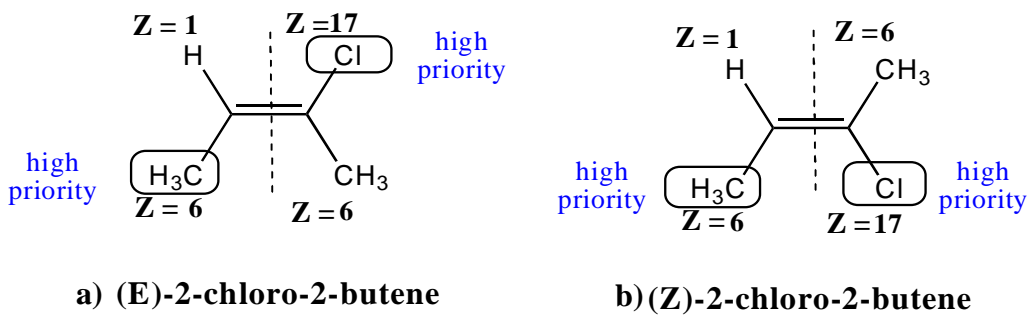

**a) (E)-2-chloro-2-butene**  **b) (Z)-2-chloro-2-butene**

**Figure 7. The two stereoisomers of the compound 2-chloro-2-butene.** The atomic number and priority of the substituents is indicated. **a)** (E)-2-chloro-2-butene: the substituents of higher priority are on opposite sides of the double bond hence the **E** assignment. **b)** (Z)-2-chloro-2-butene: the substituents of higher priority are on the same side of the double bond hence the **Z** assignment.

Following the procedure to determine the E/Z configuration of the two stereoisomers, the first step consists in determining the higher priority substituent on each end of the double bond based on its atomic number (Z) and which are indicated in the Figure 7: chlorine (Z = 17) > carbon (Z = 6) and carbon (Z = 6) > hydrogen (Z = 1). Starting with the left hand structure (Figure 7 - a)) on the left end of the double bond, the two atoms attached to the double bond are carbon and hydrogen, by the CIP priority rules, carbon has higher priority while on the right end of the double bond, the atoms attached to the double bond are carbon and chlorine, by the CIP priority rules, chlorine has higher priority. The higher priority groups are "down" and "up" respectively, therefore the substituents of higher priority are on opposite sides of the double bond, then the **E** configuration is assigned to the double bond. In the right hand structure (Figure 7 b)), the substituents of higher priority are on the same side ("down") of the double bond, then the **Z** configuration is assigned to the double bond.

Stereoisomerism can be expressed in the SMILES notation. The configuration of the atoms around double bonds is specified by the characters forward slash ("/") and back slash ("\"). These symbols represent the relative direction of the bond between two atoms, i.e., bonds that point above or below the carbon that forms the double bond. For example, the SMILES *F/C=C/Br* means that the fluorine (F) atom is below the first carbon, and the bromine (Br) atom is above the second carbon, leading to the interpretation of an **E** (*trans*) configuration. The interpretation of each directional bond is relative to the carbon atom that forms the double bond, so the sense of the symbol changes according to the order of writing, for example the compound (E)-1,2-difluoroethene may be written as F/C=C/F or C(\F)=C/F. The sequence of the atoms in the SMILES solely depends on the order of writing, and it is independent of the priorities of the atoms, therefore there is not a direct correspondence with the E/Z convention [7, 49, 50].

Stereoisomerism can also be expressed in the InChI (IUPAC International Chemical Identifier) notation presented above. The specific case of double bond stereoisomerism is encoded in the "**/b**" sub-layer of the stereochemical layer and its information is derived from the 2D (x, y) coordinates. Only explicit stereo information is included and as in SMILES, there is no simple relation between InChI and E/Z configurations since it does not represent CIP rules [8].

## 3.1 Algorithm to determine double bond stereoisomerism

The computer program that determines the double bond stereoisomerism (E/Z) of a chemical structure based on widely used linear notations as input such as SMILES or InChI was implemented in python (version 2.6) and uses OpenBabel-Pybel libraries (version 2.3.1) [51-53] which in turn use joelib2 [54] for processing chemical structures. To determine the double bond stereoisomerism according to the E/Z notation several steps are needed, namely: indentify a double bond between two carbon atoms, verify if the substituents on either end of the double bond are different, and assign the priority of each substituent on either end of the double bond according to the CIP rules. As final step in the procedure, compare the 2D coordinates of the substituents on either end of the double bond in order to determine if the substituents of higher priority are on the same or opposite sides of the double bond which allows the classification in Z or E, respectively.

The basic steps for describing a stereoisomer implemented in the algorithm are exemplified in Figure 8 for the compound (Z)-2-fluoro-3-methylpent-2-ene and detailed below.

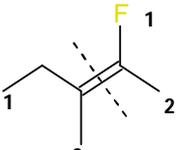

| List of SMILES representing the same structure | Canonical SMILES | Identifying a double bond and verifying if the substituents on each end of the double bond are different | Assigning the priority to each group attached to the double bond (CIP rules) | Comparing y' coordinates of the substituents after rotating the molecule | Determining the classification E-Z |
|---|---|---|---|---|---|
| C/C(F)=C(C)/CC CC/C(=C(\F)/C)/C F\C(C)=C/CC)C ..... | CC/C(=C(\F)/C)/C |  |  | y' = -0.866    y' = 0.866<br><br>y'= -1.732   y'=-9.999*10⁻⁵ | Higher priority groups are UP:<br>(1) = (1)<br>=> **Z** |
| Canonicalization Method ⇒ |  | **1** double bond<br>Left side: [CH₂-CH₃, CH₃]<br>Right side: [F, CH₃] | Left side:<br>Z = [6, 1]<br>Right side:<br>Z = [9, 6] | Left side:<br>y'(Z=6)>y'(Z=1)<br>Right side:<br>y'(Z=9) > y'(Z=6) | **(Z)-2-fluoro-3-methylpent-2-ene** |

**Figure 8. Basic steps for classifying a stereoisomer of the compound 2-fluoro-3-methylpent-2-ene using the described algorithm**: the input notation is converted in a canonical SMILES, the presence of a double bond and the substituents on each end of the double bond are verified, the priority of the substituents attached to each end of the double bond is determined according to the CIP rules, finally the double bond can be classified as E or Z comparing the y coordinates of the substituents of higher priority after rotating the molecule. In this case, the highest-priority groups on each side of the double bond are on the same side of the double bond. Fluorine is the highest priority group on the right side of the double bond, and ethyl is the highest-priority group on the left side of the molecule. This molecule can be classified as Z and the proper name is (Z)-2-fluoro-3-methylpent-2-ene.

The process of determining double bond stereoisomerism of a molecule can be detailed as follows:

**a. Identifying double bonds between two carbon atoms and verifying whether the substituents on either end of the double bond are different**

This computer program processes only molecules with double bond(s) between carbon atoms. The identification of a double bond is based on openbabel's boolean functions `IsDouble` and aromatic bonds are excluded using the function `IsAromatic`. The identification of the atoms forming the double bond is based on openbabel's boolean function `IsCarbon`.

After identifying the presence of one or more double bonds between two carbon atoms, one must verify if the substituents on either end of the double bond are different. For that purpose, the arbitrary numbering of the atoms has to be canonicalized, as mentioned above. The canonicalization method in use is the one implemented by openbabel which is based in the widely used Morgan algorithm[55] but with various improvements [51, 52]. For example, the compound (Z)-2-fluoro-3-methylpent-2-ene can be represented by the canonical SMILES CC/C(=C(\F)/C)/C, and the structure has 1 double bond (=) with different substituents on either end of the double bond: one end has the substituents $CH_2$-$CH_3$ and $CH_3$ and the other end has the substituents F and $CH_3$ (Figure 8).

**b. Assigning the priority to each substituent attached to each end of the double bond**

The substituents attached to each end of the double bond have to be numbered according to the CIP rules (detailed above) in order to determine the higher priority group. To establish the priority sequence, one must look at the atomic number of the atoms that are bonded directly to each carbon forming the double bond, arranging them in order of decreasing atomic number or, optionally, atomic mass to take into account isotopes. If two or more atoms bonded directly to the carbon atom forming the double bond have the same atomic number, then look for the first point of difference in the branch by moving out one atom (level) at a time, again comparing atomic numbers/atomic mass. This recursive method also takes into account double or triple bonds (fourth CIP rule), since higher bond orders have a smaller number of hydrogen atoms and *vice-versa*.

For the example described in Figure 8, on one end of the double bond the group with higher priority is the $CH_2$ since the atomic number of carbon is 6, and on the other side the group with higher priority is the fluorine (F) since its atomic number is 9.

**c. Compare the 2D coordinates of the substituents on either end of the double bond**

Having determined which two ligands, attached to the atoms connected by a double bond, are to be compared (higher priority), one has to calculate their relative location in relation to each other. For that purpose, one can use cartesian (x, y) structure coordinates generated from any type of molecular data structures with connectivity information. Open Babel [51, 52], version 2.3, has support for 2D coordinate generation based on the code used in the MCDL chemical structure editor [16, 61, 62]. The MCDL computer program aims to layout the molecular structure in 2D such that all bond lengths are equal. The success rate for aesthetically ideal structure layout is high, and many difficult classes of organic molecules can be depicted with confidence. Sometimes these 2D projection images have distorted (non-optimal) bond lengths and angles of structural fragments. The crucial danger, however, while processing geometrical stereoisomers comes from the fact that the reference plane can be chosen arbitrarily, and therefore, there exist different possible ways of displaying the molecule in the 2D plane. In order to find out the position of each substituent in relation to each other the 2D coordinates can be used, but since the orientation of the molecule is different in each case, one should rotate the molecule in the plane in relation to fixed axes [63, 64]. For that purpose, the first step should be the determination of the angle of rotation relative to having the double bond in an horizontal position (Equation **1**). The calculation of the angle of rotation (**θ**) is based on the arc tangent function of the slope of the line formed by the double bond in relation to the horizontal line, that is determined using the coordinates of the two carbon atoms forming the double bond (($x_0$, $y_0$), ($x_1$, $y_1$)).

$$\boldsymbol{\theta = arctan\frac{y_1 - y_0}{x_1 - x_0}} \quad (1)$$

A vertical line has an undefined slope and therefore this case will be determined using another procedure, explained below.

To rotate the molecule anti-clockwise around the origin by **θ** is equivalent to replacing every point with coordinates (x,y) by the point with coordinates (x',y'), where:

$$\begin{bmatrix} x' \\ y' \end{bmatrix} = \begin{bmatrix} cos\theta & -sin\theta \\ sin\theta & cos\theta \end{bmatrix} \begin{bmatrix} x \\ y \end{bmatrix} \Leftrightarrow \begin{matrix} x' = xcos\theta - ysin\theta \\ y' = xsin\theta + ycos\theta \end{matrix} \quad (2)$$

Similarly, for a clockwise rotation around the origin by **θ**:

$$\begin{bmatrix} x' \\ y' \end{bmatrix} = \begin{bmatrix} cos\theta & sin\theta \\ -sin\theta & cos\theta \end{bmatrix} \begin{bmatrix} x \\ y \end{bmatrix} \Leftrightarrow \begin{matrix} x' = xcos\theta + ysin\theta \\ y' = -xsin\theta + ycos\theta \end{matrix} \quad (3)$$

It is also possible to use the equation **(2)** for clockwise rotation using the opposite angle (**-θ**). If the slope of the line formed by the double bond is negative, the molecule should be rotated anti-clockwise and therefore one should use equation **(2)**. If the slope of the line formed by the double bond is positive, the molecule should be rotated clockwise and therefore one should use equation **(3)** or **(2)** with $\theta = -\theta$.

Using the rotated coordinates (x', y') is now possible to determine the substituents position in relation to each other, comparing the y' coordinates. The substituent with higher y' is up (the value 1 is assigned) while the substituent with lower y' is down (the value -1 is assigned).

For the molecules with vertical orientation (undefined slope) the procedure can be simplified to the comparison of the un-rotated x coordinates. The substituent with higher x is up (the value 1 is assigned) while the substituent with lower x is down (the value -1 is assigned).

Comparing the y' coordinates of the substituents on each side of the double bond for the rotated structure represented in Figure 8, leads to the conclusion that the substituents of higher priority have higher y' and therefore both are classified as 1.

### d. Determining the classification E-Z using the 2D coordinates

According to the comparison of 2D coordinates of the substituents on either end of the double bond determined in **c** and comparing the two sides of the double bond, if the two groups with the higher priorities are on the same side of the double bond ((-1)(-1) or (1)(1)), that is described as a **Z**-double bond (the value 1 is assigned) otherwise if the two groups with the higher priorities are on opposite sides of the double bond ((-1)(1) or (1)(-1)), that is described as a **E**-double bond (the value -1 is assigned). For the example described in Figure 8, the substituents of higher priority were both classified as 1, hence the double bond is described as Z.

## 3.2 Detailed example of the steps of execution of the double bond stereoisomerism classification program

In order to analyze in a more detailed way the execution of the developed program, another example, (Z)-1-bromo-3-(chloromethyl)-6-methyl-4-propylhept-3-ene (Figure 9), will be described together with the program's outcomes in each step.

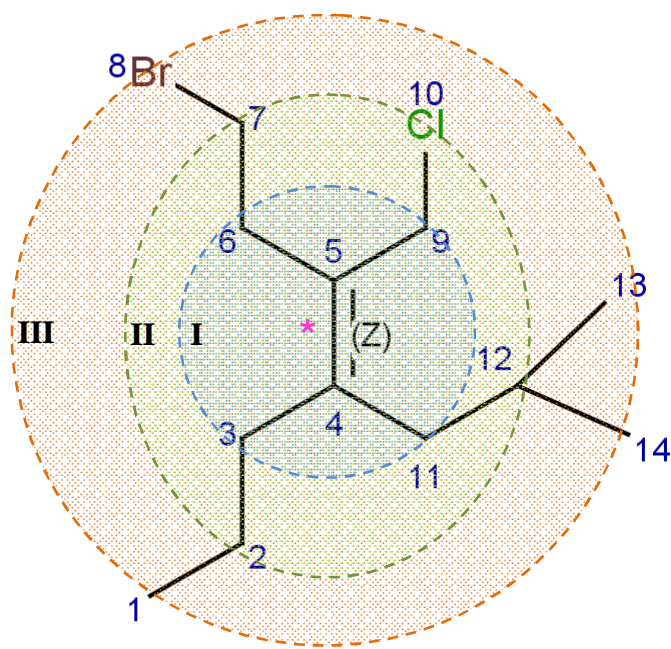

**Figure 9. The compound (Z)-1-bromo-3-(chloromethyl)-6-methyl-4-propylhept-3-ene (CCC/C(=C(\CCBr)/CCl)/CC(C)C).** The double bond is marked with an asterisk (between the carbon atoms with canonical numbers 4 and 5) and the canonical numbers of each atom are indicated near each atom symbol. The hydrogen atoms were made implicit for readability purposes (canonical numbers between 15 and 36). To apply outward exploration of groups, starting from the carbon atoms forming the double bond, the atoms are divided in levels, which are indicated in the figure with colored shells: **I)** level 1, marked with a blue shell, containing the atoms with the canonical numbers 3, 6, 11 and 9; **II)** level 2, marked with a green shell, containing the atoms with the canonical numbers 2, 7, 10 and 12 and the non-represented hydrogen atoms 20,

21, 22, 23, 26, 27, 28 and 29; **III)** level 3, marked with a orange shell, containing the atoms with the canonical numbers 1, 8, 13 and 14 and the non-represented hydrogen atoms 18, 19, 24, 25 and 30; **IV)** level 4, not represented in the figure, containing the non-represented hydrogen atoms 15, 16, 17, 31, 32, 33, 34, 35 and 36. The atomic numbers (Z) of the atoms represented in the figure are: bromine (Br) (Z = 35), chlorine (Cl) (Z = 17), carbon (C) (Z = 6), hydrogen (H) (Z = 1).

The canonicalization of this structure generates the following SMILES, CCC/C(=C(\CCBr)/CCl)/CC(C)C. Figure 9 represents the structure of (Z)-1-bromo-3-(chloromethyl)-6-methyl-4-propylhept-3-ene with implicit hydrogen atoms for readability purposes (canonical numbers between 15 and 36) and the represented atoms numbered according to the canonical order. Thus, a simple 6-steps process is followed to determine its double bond stereosiomerism:

**1.** After identifying that the compound has at least one double bond (that is not aromatic) using openbabel, it is converted to a 2D MDL MOLFile using openbabel or Marvin molconverter to make available 2D coordinates of its atoms (Figure 10).

```
a) OpenBabel 10291214122D

36 35  0  0  0  0  0  0  0  0999 V2000
   -4.0000   -0.0000    0.0000 C   0  0  0  0  0  0  0  0  0  0  0  0
   -4.5000   -0.8660    0.0000 C   0  0  0  0  0  0  0  0  0  0  0  0
   -4.0000   -1.7321    0.0000 C   0  0  0  0  0  0  0  0  0  0  0  0
   -3.0000   -1.7321    0.0000 C   0  0  0  0  0  0  0  0  0  0  0  0
   -2.5000   -2.5981    0.0000 C   0  0  0  0  0  0  0  0  0  0  0  0
   -3.0000   -3.4641    0.0000 C   0  0  0  0  0  0  0  0  0  0  0  0
   -4.0000   -3.4641    0.0000 C   0  0  0  0  0  0  0  0  0  0  0  0
   -4.8660   -3.9641    0.0000 Br  0  0  0  0  0  0  0  0  0  0  0  0
   -1.5000   -2.5981    0.0000 C   0  0  0  0  0  0  0  0  0  0  0  0
   -0.6340   -3.0981    0.0000 Cl  0  0  0  0  0  0  0  0  0  0  0  0
   -2.5000   -0.8660    0.0000 C   0  0  0  0  0  0  0  0  0  0  0  0
   -1.5000   -0.8660    0.0000 C   0  0  0  0  0  0  0  0  0  0  0  0
   -1.5000    0.1340    0.0000 C   0  0  0  0  0  0  0  0  0  0  0  0
   -0.5000   -0.8660    0.0000 C   0  0  0  0  0  0  0  0  0  0  0  0
   -4.0000    1.0000    0.0000 H   0  0  0  0  0  0  0  0  0  0  0  0
   -4.8660    0.5000    0.0000 H   0  0  0  0  0  0  0  0  0  0  0  0
   -3.0000   -0.0000    0.0000 H   0  0  0  0  0  0  0  0  0  0  0  0
   -5.3660   -1.3660    0.0000 H   0  0  0  0  0  0  0  0  0  0  0  0
   -5.3660   -0.3660    0.0000 H   0  0  0  0  0  0  0  0  0  0  0  0
   -4.0000   -2.7321    0.0000 H   0  0  0  0  0  0  0  0  0  0  0  0
   -4.8660   -2.2321    0.0000 H   0  0  0  0  0  0  0  0  0  0  0  0
   -3.0000   -4.4641    0.0000 H   0  0  0  0  0  0  0  0  0  0  0  0
   -2.1340   -3.9641    0.0000 H   0  0  0  0  0  0  0  0  0  0  0  0
   -4.0000   -4.4641    0.0000 H   0  0  0  0  0  0  0  0  0  0  0  0
   -4.5000   -2.5981    0.0000 H   0  0  0  0  0  0  0  0  0  0  0  0
   -1.5000   -3.5981    0.0000 H   0  0  0  0  0  0  0  0  0  0  0  0
   -1.0000   -1.7321    0.0000 H   0  0  0  0  0  0  0  0  0  0  0  0
   -2.5000    0.1340    0.0000 H   0  0  0  0  0  0  0  0  0  0  0  0
   -3.3660   -0.3660    0.0000 H   0  0  0  0  0  0  0  0  0  0  0  0
   -1.5000   -1.8660    0.0000 H   0  0  0  0  0  0  0  0  0  0  0  0
   -1.0000    1.0000    0.0000 H   0  0  0  0  0  0  0  0  0  0  0  0
   -0.5000    0.1340    0.0000 H   0  0  0  0  0  0  0  0  0  0  0  0
   -2.3660    0.6340    0.0000 H   0  0  0  0  0  0  0  0  0  0  0  0
    0.3660   -1.3660    0.0000 H   0  0  0  0  0  0  0  0  0  0  0  0
   -0.5000   -1.8660    0.0000 H   0  0  0  0  0  0  0  0  0  0  0  0
    0.0000    0.0000    0.0000 H   0  0  0  0  0  0  0  0  0  0  0  0

b)
  1  2  1  0  0  0  0
  1 15  1  0  0  0  0
  1 16  1  0  0  0  0
  1 17  1  0  0  0  0
  2  3  1  0  0  0  0
  2 18  1  0  0  0  0
  2 19  1  0  0  0  0
  3  4  1  0  0  0  0
  3 20  1  0  0  0  0
  3 21  1  0  0  0  0
  4  5  2  0  0  0  0
  4 11  1  0  0  0  0
  5  6  1  0  0  0  0
  5  9  1  0  0  0  0
  6  7  1  0  0  0  0
  6 22  1  0  0  0  0
  6 23  1  0  0  0  0
  7  8  1  0  0  0  0
  7 24  1  0  0  0  0
  7 25  1  0  0  0  0
  9 10  1  0  0  0  0
  9 26  1  0  0  0  0
  9 27  1  0  0  0  0
 11 12  1  0  0  0  0
 11 28  1  0  0  0  0
 11 29  1  0  0  0  0
 12 13  1  0  0  0  0
 12 14  1  0  0  0  0
 12 30  1  0  0  0  0
 13 31  1  0  0  0  0
 13 32  1  0  0  0  0
 13 33  1  0  0  0  0
 14 34  1  0  0  0  0
 14 35  1  0  0  0  0
 14 36  1  0  0  0  0
M  END
```

**Figure 10. 2D MDL MOLFile for the compound (Z)-1-bromo-3-(chloromethyl)-6-methyl-4-propylhept-3-ene generated by openbabel**: **a)** 2D coordinates of the atoms in the molecule; **b)** connection table of the atoms in the molecule and corresponding type of bonds.

**2.** A recursive process is used to retrieve the atomic numbers of each atom directly attached to each end of the double bond and each branch is traversed recursively. Four lists are created (two starting from each carbon of the double bond), each of them starting with the atomic number (Z) of the atom directly attached to the carbon atoms forming the double bond (all carbons (C) (indexes = 3, 11, 6, 9 , Z = 6), followed by lists representing a sequential topological distance to the carbon atom forming the double bond (levels) with two-atoms tuples (canonical indexes) indicating the connectivity of each branch:

Branches starting from the carbon of the double bond with canonical number 4:

[6, [(3, 20), (3, 2), (3, 21)], [(2, 1), (2, 18), (2, 19)], [(1, 17), (1, 15), (1, 16)]]
[6, [(11, 12), (11, 28), (11, 29)], [(12, 13), (12, 14), (12, 30)], [(13, 33), (13, 31), (13, 32), (14, 34), (14, 35), (14, 36)]]

Branches starting from the carbon of the double bond with canonical number 5:

[6, [(6, 23), (6, 22), (6, 7)], [(7, 25), (7, 24), (7, 8)]]
[6, [(9, 10), (9, 27), (9, 26)]]

**3.** The atoms of each branch are sorted in descending order of their topological distances to the carbon of the double bond (levels) and inside each level in descending order according the atomic numbers. Each branch is represented by a tuple and the connectivity tuples are replaced by lists of atomic numbers (ordered in descending order) of the atoms in each level and the last position of each tuple represents the branch number on each side of the double bond, according to the canonical order (0 or 1):

Branches starting from the carbon of the double bond with canonical number 4:

([[6], [6, 1, 1], [6, 1, 1], [1, 1, 1]], 0)
([[6], [6, 1, 1], [6, 6, 1], [1, 1, 1, 1, 1, 1]], 1)

Branches starting from the carbon of the double bond with canonical number 5:

([[6], [6, 1, 1], [35, 1, 1]], 0)
([[6], [17, 1, 1]], 1)

**4.** The lists on each side of the double bond are then compared between themselves, if they are equal then the branches are the same and there is no double bond stereoisomerism, else the procedure continues to retrieve their priority order, as per the CIP rules. This involves solely comparing each list in the two tuples on each side of the double bond to the others to sort the branches according to their atom content:

Branches starting from the carbon of the double bond with canonical number 4:
([[6], [6, 1, 1], [6, 6, 1], [1, 1, 1, 1, 1, 1]], 1)
([[6], [6, 1, 1], [6, 1, 1], [1, 1, 1]], 0)

Branches starting from the carbon of the double bond with canonical number 5:
([[6], [17, 1, 1]], 1)
([[6], [6, 1, 1], [35, 1, 1]], 0)

**5.** The first branch of each list (two sides of the double bond) is the higher priority one, now it is needed to determinate their location in relation to each other. For that purpose several sub-steps are needed:

**a.** Calculate the slope of the double bond (using the carbon atoms (x, y) coordinates:

Carbon with canonical number 4: (x, y) = (-3.0, -1.732)
Carbon with canonical number 5: (x, y) = (-2.5, -2.598)
Slope of the double bond: -1.732

**b.** Calculate the angle of rotation in radians ($\theta$) based on the slope of the double bond:
$$\theta = -1.0472$$

**c.** The slope of the double bond is negative, thus the molecule should be rotated anti-clockwise. Using the coordinates (x, y) of each group directly attached to the carbon atoms forming the double bond, calculate the rotated coordinates. The y rotated coordinates (y') should be used to compare the position of the branches in each side of the double bond. The first branch on each side of the double bond is the higher priority one.

Branches starting from the carbon of the double bond with canonical number 4:

**atom 11 (x, y) = (-2.5, -0.866) → y' (atom 11) = 1.732**
**atom 3 (x, y) = (-4.0, -1.732) → y' (atom 3) = 2.598**

The y' coordinate of the first atom (canonical number 11) is lower than the y' coordinate of the second (canonical number 3), therefore the higher priority group is down and the value **-1** is assigned.

Branches starting from the carbon of the double bond with canonical number 5:

**atom 9 (x, y) = (-1.5, -2.598) → y' (atom 9) = -5.000 * 10$^{-5}$**
**atom 6 (x, y) = (-3.0, -3.464) → y' (atom 6) = 0.866**

The y' coordinate of the first atom (canonical number 9) is lower than the y' coordinate of the second (canonical number 6), therefore the higher priority group is down and the value **-1** is assigned.

6. Finally, it is possible to determine the classification in E-Z based on the relative position of the high priority substituents directly attached to each side of the double bond:

branch of higher priority connected to the carbon 4 is down (**-1**) * branch of higher priority connected to the carbon 5 is down (**-1**) = **1,** which corresponds to a **Z**-double bond (1).

The worst-case time complexity of an algorithm gives an idea of the number of operations needed to solve it given any input of size *n* [58]. In the case of the given computer program, the most complex algorithm is the recursive search for the ligands attached to each double bond, in the worst-case scenario for each double bond it goes through all atoms in the molecule. The complexity for the worst-case scenario (using the big O notation) of the given algorithm, based on well-established algorithms and their known complexity [59], is *O(D\*A\*log(A))*, where *D* is the number of double bonds and *A* is the number of atoms in the molecule.

The development of this computer program was guided and tested by calculating stereoisomerism around double bonds to a set of 220 molecules with this type of stereochemistry. For this set, there were 2 failures in the correct stereoisomerism classification: (E,E,Z)-1,5,9-Cyclododecatriene and [16]Annulene. It was further verified that for the first compound ((E,E,Z)-1,5,9-Cyclododecatriene) the problem was in the openbabel's canonicalization, since stereochemical information is lost during this process as shown in Figure 11. The exact same problem was found for another 1,5,9-Cyclododecatriene stereoisomer, the (E,E,E)-1,5,9-Cyclododecatriene, its original SMILES C1=C/CC/C=C/CC/C=C/CC/1 is canonicalized in C1CC=CCCC=CCCC=C1, nevertheless the output classification is correct, because the coordinates thus allow. Openbabel's implementation of the canonical algorithm was tested using more than 5 millions of molecules in the eMolecules [65] catalog (which does not contain the stereoisomers of 1,5,9-Cyclododecatriene) and it produced less than 0.001% errors, thus this implementation performance can be considered good [51].

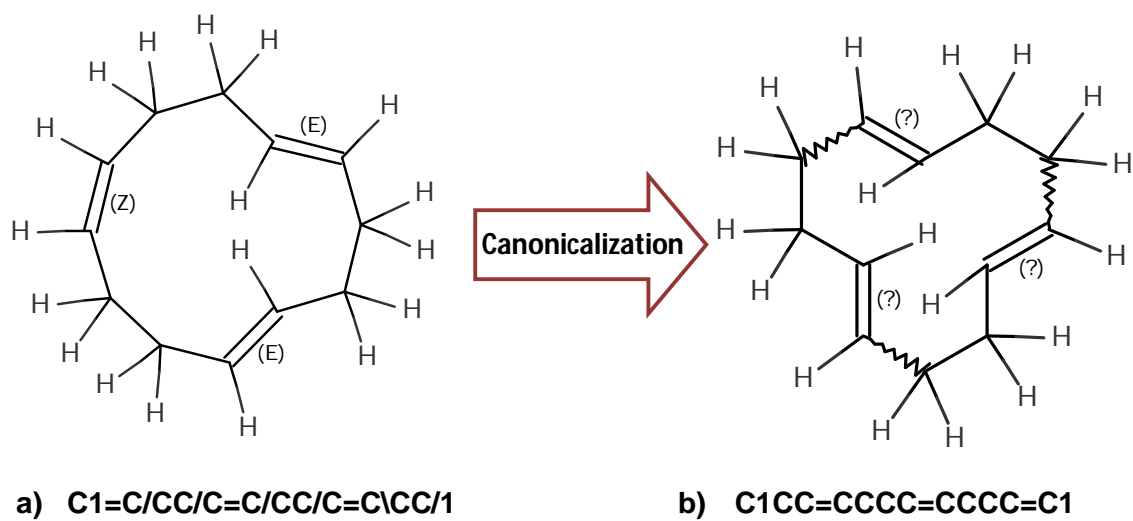

a)  C1=C/CC/C=C/CC/C=C\CC/1          b)  C1CC=CCCC=CCCC=C1

**Figure 11. Canonicalization of the SMILES for the compound (E,E,Z)-1,5,9-Cyclododecatriene using openbabel**: a) represents the molecular structure using the original SMILES; b) represents the molecular structure using the canonical SMILES.

For the compound [16]Annulene the problem is different, the 2D coordinates of the hydrogen atoms are not correct whereas, many of them have exactly the same location in the plane causing an incorrect classification of E-Z stereoisomerism. The same SMILES was used to generate the 2D coordinates using a command line tool, molconverter, provided by another software package, Marvin Beans[18], and the obtained coordinates allow a correct classification of the E-Z stereoisomerism. Therefore, it was included an optional parameter in the program that allows the user to select the implementation that will generate the 2D coordinates: openbabel or molconverter - Marvin Beans.

## 4. Web Tool

With the developed algorithm, a public and free Web tool (*Chirality and Double Bond Stereoisomerism Calculator*) has been implemented. The objective was to produce an application simple to use with easily readable results that would allow the determination of chirality and double bond stereoisomerism for a list of molecules. The user can input a list of molecules using their common name, SMILES string, or InChI identifier. The common name is resolved using the Chemical Identifier Resolver (http://cactus.nci.nih.gov/chemical/structure), directly called by the application. The output produced consists in the identifiers of the molecule, a graphical representation of the molecule with the canonical numeration of atoms (generated using openbabel) and the algorithm output: the position of the stereocenters or double bonds with stereosiomerism and its classification according to the CIP rules (Figure 12).

**Figure 12.** *Chirality and Double Bond Stereoisomerism Calculator* (http://nams.lasige.di.fc.ul.pt/tools.php) **example, showing the output for running a sample list of molecules.**

This Web tool was mainly developed in the PHP programming language. The application communicates with a Python script that uses openbabel for converting the different representations of the molecule and determining the chirality and double bond stereoisomerism of each structure according with the described algorithm.

The *Chirality and Double Bond Stereoisomerism Calculator* is available at http://nams.lasige.di.fc.ul.pt/tools.php

## 5. Summary

This section provides a brief summary of the full report with an emphasis on the report's major findings, recommendations, and conclusions.

- The algorithm described in this report deals with the automatic detection and classification of stereoisomers: stereochemistry of a chiral carbon atom and stereochemistry of a double bond.

- The purpose of this work was to provide a binary classification of the isomer type according to CIP rules while using as input widely used line notations such as SMILES or InChI, in order to be able to compare molecules taking this point into account.

- Most of the existing software packages are proprietary and/or commercial, and they do not allow batch processing of molecules encoded as line representations such as SMILES or InChI. Therefore it was not possible to perform a comparative study in terms of efficiency and complexity.

- The algorithm was successfully tested on a sample of 3000 compounds of which 210 contain explicit information concerning chirality and 220 have explicit information about double bond stereoisomerism. The issues found were described and addressed.

- Future developments in the algorithm intend to include the detection and classification of other types of stereochemistry and the fifth CIP rule.

**Acknowledgements**

ALT gratefully acknowledges Fundação para a Ciência e a Tecnologia for a doctoral grant (SFRH/BD/64487/2009).